\documentclass[longauth]{aa}
\usepackage{graphicx}
\usepackage{txfonts}
\usepackage{xcolor}
\usepackage{hyperref}
\hypersetup{colorlinks, linkcolor=blue, citecolor=blue, urlcolor=blue}

\newcommand{\pivec}{\mbox{\boldmath $\pi$}}
\newcommand{\muvec}{\mbox{\boldmath $\mu$}}

\newcommand{\te}{t_{\rm E}}
\newcommand{\thetae}{\theta_{\rm E}}

\newcommand{\pie}{\pi_{\rm E}}

\newcommand{\dl}{D_{\rm L}}
\newcommand{\ds}{D_{\rm S}}

\definecolor{brown}{rgb}{0.59, 0.29, 0.0}
\definecolor{darkgreen}{rgb}{0.0, 0.42, 0.24}
\definecolor{darkblue}{rgb}{0.01, 0.31, 0.59}
\definecolor{darkblue}{rgb}{0.0, 0.25, 0.42}
\definecolor{blue}{rgb}{0.0,0.0,1.0}
\definecolor{green}{rgb}{0.0,1.0,0.0}

\begin{document}

\title{Four microlensing giant planets detected through signals produced by
minor-image perturbations}
\titlerunning{Microlensing giant planets detected through signals produced by
minor-image perturbations}

\author{
     Cheongho~Han\inst{\ref{inst1}} 
\and Ian~A.~Bond\inst{\ref{inst2}}
\and Chung-Uk~Lee\inst{\ref{inst3},\ref{inst37}}
\and Andrew~Gould\inst{\ref{inst4},\ref{inst5}}      
\\
(Leading authors)
\\
     Michael~D.~Albrow\inst{\ref{inst6}}   
\and Sun-Ju~Chung\inst{\ref{inst3}}      
\and Kyu-Ha~Hwang\inst{\ref{inst3}} 
\and Youn~Kil~Jung\inst{\ref{inst3}} 
\and Yoon-Hyun~Ryu\inst{\ref{inst3}} 
\and Yossi~Shvartzvald\inst{\ref{inst7}}   
\and In-Gu~Shin\inst{\ref{inst8}} 
\and Jennifer~C.~Yee\inst{\ref{inst8}}   
\and Hongjing~Yang\inst{\ref{inst9}}     
\and Weicheng~Zang\inst{\ref{inst8},\ref{inst9}}     
\and Sang-Mok~Cha\inst{\ref{inst3},\ref{inst10}} 
\and Doeon~Kim\inst{\ref{inst1}}
\and Dong-Jin~Kim\inst{\ref{inst3}} 
\and Seung-Lee~Kim\inst{\ref{inst3}} 
\and Dong-Joo~Lee\inst{\ref{inst3}} 
\and Yongseok~Lee\inst{\ref{inst3},\ref{inst10}} 
\and Byeong-Gon~Park\inst{\ref{inst3}} 
\and Richard~W.~Pogge\inst{\ref{inst5}}
\\
(The KMTNet Collaboration)
\\
     Fumio~Abe\inst{\ref{inst11}}
\and Ken~Bando\inst{\ref{inst12}}
\and Richard~Barry\inst{\ref{inst13}}
\and David~P.~Bennett\inst{\ref{inst13},\ref{inst14}}
\and Aparna~Bhattacharya\inst{\ref{inst13},\ref{inst14}}
\and Hirosame~Fujii\inst{\ref{inst11}}
\and Akihiko~Fukui\inst{\ref{inst15},}\inst{\ref{inst16}}
\and Ryusei~Hamada\inst{\ref{inst12}}
\and Shunya~Hamada\inst{\ref{inst12}}
\and Naoto~Hamasaki\inst{\ref{inst12}}
\and Yuki~Hirao\inst{\ref{inst17}}
\and Stela~Ishitani Silva\inst{\ref{inst13},\ref{inst18}}
\and Yoshitaka~Itow\inst{\ref{inst11}}
\and Rintaro~Kirikawa\inst{\ref{inst12}}
\and Naoki~Koshimoto\inst{\ref{inst12}}
\and Yutaka~Matsubara\inst{\ref{inst11}}
\and Shota~Miyazaki\inst{\ref{inst19}}
\and Yasushi~Muraki\inst{\ref{inst11}}
\and Tutumi~Nagai\inst{\ref{inst12}}
\and Kansuke~Nunota\inst{\ref{inst12}}
\and Greg~Olmschenk\inst{\ref{inst13}}
\and Cl{\'e}ment~Ranc\inst{\ref{inst20}}
\and Nicholas~J.~Rattenbury\inst{\ref{inst21}}
\and Yuki~Satoh\inst{\ref{inst12}}
\and Takahiro~Sumi\inst{\ref{inst12}}
\and Daisuke~Suzuki\inst{\ref{inst12}}
\and Mio~Tomoyoshi\inst{\ref{inst12}}
\and Paul~J.~Tristram\inst{\ref{inst22}}
\and Aikaterini~Vandorou\inst{\ref{inst13},\ref{inst14}}
\and Hibiki~Yama\inst{\ref{inst12}}
\and Kansuke~Yamashita\inst{\ref{inst12}}
\\
(The MOA Collaboration)\\
     Etienne~Bachelet\inst{\ref{inst23}}
\and Paolo~Rota\inst{\ref{inst24},\ref{inst25}}
\and Valerio~Bozza\inst{\ref{inst24},\ref{inst25}}
\and Pawe{\l}~Zielinski\inst{\ref{inst26}}
\and Rachel~A.~Street\inst{\ref{inst27}}
\and Yiannis~Tsapras\inst{\ref{inst28}}
\and Markus~Hundertmark\inst{\ref{inst28}}
\and Joachim~Wambsganss\inst{\ref{inst28}}
\and {\L}ukasz~Wyrzykowski\inst{\ref{inst29}}
\and Roberto~Figuera~Jaimes\inst{\ref{inst30},\ref{inst31}}
\and Arnaud~Cassan\inst{\ref{inst32}}
\and Martin~Dominik\inst{\ref{inst33}}
\and Krzysztof~A.~Rybicki\inst{\ref{inst34},\ref{inst35}}
\and Markus~Rabus\inst{\ref{inst36}}
\\
(The OMEGA collaboration)
}

\institute{
      Department of Physics, Chungbuk National University, Cheongju 28644, Republic of Korea\label{inst1}                                                                                                         
\and  Institute of Natural and Mathematical Science, Massey University, Auckland 0745, New Zealand\label{inst2}                                                                                                   
\and  Korea Astronomy and Space Science Institute, Daejon 34055, Republic of Korea\label{inst3}                                                                                                                   
\and  Max Planck Institute for Astronomy, K\"onigstuhl 17, D-69117 Heidelberg, Germany\label{inst4}                                                                                                               
\and  Department of Astronomy, The Ohio State University, 140 W. 18th Ave., Columbus, OH 43210, USA\label{inst5}                                                                                                  
\and  University of Canterbury, Department of Physics and Astronomy, Private Bag 4800, Christchurch 8020, New Zealand\label{inst6}                                                                                
\and  Department of Particle Physics and Astrophysics, Weizmann Institute of Science, Rehovot 76100, Israel\label{inst7}                                                                                          
\and  Center for Astrophysics $|$ Harvard \& Smithsonian 60 Garden St., Cambridge, MA 02138, USA\label{inst8}                                                                                                     
\and  Department of Astronomy and Tsinghua Centre for Astrophysics, Tsinghua University, Beijing 100084, China\label{inst9}                                                                                       
\and  School of Space Research, Kyung Hee University, Yongin, Kyeonggi 17104, Republic of Korea\label{inst10}                                                                                                     
\and  Institute for Space-Earth Environmental Research, Nagoya University, Nagoya 464-8601, Japan\label{inst11}                                                                                                   
\and  Department of Earth and Space Science, Graduate School of Science, Osaka University, Toyonaka, Osaka 560-0043, Japan\label{inst12}                                                                          
\and  Code 667, NASA Goddard Space Flight Center, Greenbelt, MD 20771, USA\label{inst13}                                                                                                                          
\and  Department of Astronomy, University of Maryland, College Park, MD 20742, USA\label{inst14}                                                                                                                  
\and  Department of Earth and Planetary Science, Graduate School of Science, The University of Tokyo, 7-3-1 Hongo, Bunkyo-ku, Tokyo 113-0033, Japan\label{inst15}                                                 
\and  Instituto de Astrof{\'i}sica de Canarias, V{\'i}a L{\'a}ctea s/n, E-38205 La Laguna, Tenerife, Spain\label{inst16}                                                                                          
\and  Institute of Astronomy, Graduate School of Science, The University of Tokyo, 2-21-1 Osawa, Mitaka, Tokyo 181-0015, Japan\label{inst17}                                                                      
\and  Oak Ridge Associated Universities, Oak Ridge, TN 37830, USA\label{inst18}                                                                                                                                   
\and  Institute of Space and Astronautical Science, Japan Aerospace Exploration Agency, 3-1-1 Yoshinodai, Chuo, Sagamihara, Kanagawa 252-5210, Japan\label{inst19}                                                
\and  Sorbonne Universit\'e, CNRS, UMR 7095, Institut d'Astrophysique de Paris, 98 bis bd Arago, 75014 Paris, France\label{inst20}                                                                                
\and  Department of Physics, University of Auckland, Private Bag 92019, Auckland, New Zealand\label{inst21}                                                                                                       
\and  University of Canterbury Mt.~John Observatory, P.O. Box 56, Lake Tekapo 8770, New Zealand\label{inst22}                                                                                                     
\and  IPAC, Mail Code 100-22, Caltech, 1200 E. California Blvd., Pasadena, CA 91125, USA\label{inst23}                                                                                                            
\and  Dipartimento di Fisica ``E.R. Caianiello'', Universit\`a di Salerno, Via Giovanni Paolo II 132, 84084 Fisciano, Italy\label{inst24}                                                                         
\and  Instituto Nazionale di Fisica Nucleare, Sezione di Napoli, Via Cintia, 80126 Napoli, Italy\label{inst25}                                                                                                    
\and  Institute of Astronomy, Faculty of Physics, Astronomy and Informatics, Nicolaus Copernicus University in Toru{\'n}, Grudzi\k{a}dzka 5, 87-100 Toru{\'n}, Poland\label{inst26}                               
\and  Las Cumbres Observatory Global Telescope Network, Inc., 6740 Cortona Drive, Suite 102, Goleta, CA 93117, USA\label{inst27}                                                                                  
\and  Zentrum f{\"u}r Astronomie der Universit{\"a}t Heidelberg, Astronomisches Rechen-Institut, M{\"o}nchhofstr. 12-14, 69120 Heidelberg, Germany\label{inst28}                                                  
\and  Astronomical Observatory, University of Warsaw, Al.~Ujazdowskie~4, 00-478~Warszawa, Poland\label{inst29}                                                                                                    
\and  Millennium Institute of Astrophysics MAS, Nuncio Monsenor Sotero Sanz 100, Of. 104, Providencia, Santiago, Chile\label{inst30}                                                                              
\and  Instituto de Astrof\'isica, Facultad de F\'isica, Pontificia Universidad Cat\'olica de Chile, Av. Vicu\~na Mackenna 4860, 7820436 Macul, Santiago, Chile\label{inst31}                                      
\and  Institut d'Astrophysique de Paris, Sorbonne Universit\'e, CNRS, UMR 7095, 98 bis bd Arago, F-75014 Paris, France\label{inst32}                                                                              
\and  University of St Andrews, Centre for Exoplanet Science, School of Physics \& Astronomy, North Haugh, St Andrews, KY16 9SS, United Kingdom\label{inst33}                                                     
\and  Astronomical Observatory, University of Warsaw, Al.~Ujazdowskie~4, 00-478~Warszawa, Poland\label{inst34}                                                                                                    
\and  Department of Particle Physics and Astrophysics, Weizmann Institute of Science, Rehovot 76100, Israel\label{inst35}                                                                                         
\and  Departamento de Matem{\'a}tica y F{\'i}sica Aplicadas, Facultad de Ingenier{\'i}a, Universidad Cat{\'o}lica de la Sant{\'i}sima Concepci{\'o}n, Alonso de Rivera 2850, Concepci{\'o}n, Chile\label{inst36}  
\and  Corresponding author\label{inst37}                                                                                                                                                                          
}                                                                                                                                                                                           
\date{Received ; accepted}                                                                                                                                                                                      
                                                                                                                                                                            

\abstract
{}
{
We investigated the nature of the anomalies appearing in four microlensing events
KMT-2020-BLG-0757, KMT-2022-BLG-0732, KMT-2022-BLG-1787, and KMT-2022-BLG-1852. 
The light curves of these events commonly exhibit initial bumps followed by 
subsequent troughs that extend across a substantial portion of the light curves.

}
{
We performed thorough modeling of the anomalies to elucidate their characteristics. 
Despite their prolonged durations, which differ from the usual brief anomalies 
observed in typical planetary events, our analysis revealed that each anomaly in 
these events originated from a planetary companion located within the Einstein 
ring of the primary star. It was found that the initial bump arouse when the source 
star crossed one of the planetary caustics, while the subsequent trough feature
occurred as the source traversed the region of minor image perturbations lying between 
the pair of planetary caustics.
}
{
The estimated masses of the host and planet, their mass ratios, and the distance to the 
discovered planetary systems are
$(M_{\rm host}/M_\odot, M_{\rm planet}/M_{\rm J}, q/10^{-3}, \dl/{\rm kpc}) =
 (0.58^{+0.33}_{-0.30}, 10.71^{+6.17}_{-5.61}, 17.61\pm 2.25,6.67^{+0.93}_{-1.30})$ for KMT-2020-BLG-0757,
$(0.53^{+0.31}_{-0.31}, 1.12^{+0.65}_{-0.65},  2.01 \pm 0.07, 6.66^{+1.19}_{-1.84})$ for KMT-2022-BLG-0732,
$(0.42^{+0.32}_{-0.23}, 6.64^{+4.98}_{-3.64}, 15.07\pm 0.86, 7.55^{+0.89}_{-1.30})$ for KMT-2022-BLG-1787, and
$(0.32^{+0.34}_{-0.19}, 4.98^{+5.42}_{-2.94},  8.74\pm 0.49, 6.27^{+0.90}_{-1.15})$ for KMT-2022-BLG-1852.
These parameters indicate that all the planets are giants with masses exceeding the mass of Jupiter 
in our solar system and the hosts are low-mass stars with masses substantially less massive than 
the Sun.
}
{}


\maketitle

\begin{table*}[t]
\small
\caption{Coordinates, baseline magnitude, and extinction.  \label{table:one}}
\begin{tabular}{lllll}
\hline\hline
\multicolumn{1}{c}{Event}                      &
\multicolumn{1}{c}{(RA, DEC)$_{\rm J2000}$}    &
\multicolumn{1}{c}{$(l,b)$}                    &
\multicolumn{1}{c}{$I_{\rm base}$}             &
\multicolumn{1}{l}{$A_I$}                     \\
\hline
KMT-2020-BLG-0757 & (18:04:32.48, -27:55:21.68) & $(+2^\circ\hskip-2pt .9746, -3^\circ\hskip-2pt .0803)$  &  18.93 &  1.11  \\
KMT-2022-BLG-0732 & (17:40:27.37, -35:51:27.97) & $(-6^\circ\hskip-2pt .4644, -2^\circ\hskip-2pt .7086)$  &  17.82 &  2.54  \\
KMT-2022-BLG-1787 & (17:50:22.37, -30:23:58.88) & $(-0^\circ\hskip-2pt .7198, -1^\circ\hskip-2pt .6400)$  &  18.45 &  2.84  \\
KMT-2022-BLG-1852 & (18:19:10.92, -24:16:16.90) & $(+7^\circ\hskip-2pt .7621, -4^\circ\hskip-2pt .2222)$  &  18.72 &  0.77  \\
\hline
\end{tabular}
\end{table*}

\section{Introduction} \label{sec:one}

In contrast to the simplified representation of a planetary microlensing signal as a brief,
discontinuous deviation in the smooth lensing light curve caused by the host star of the 
planet, the manifestations of planets exhibit a high degree of variability. These signals 
arise when the source crosses or approaches the caustic generated by the presence of a 
planet \citep{Mao1991, Gould1992}.  Caustics in microlensing represent the positions at 
which the magnification of a point source becomes infinitely large. Caustics induced by 
planets form single or multiple closed curves. The characteristics of these caustic curves, 
including their number, location, and size, vary depending on the separation and mass ratio 
between the planet and its host star.  Combined with the varied trajectories of source stars, 
planetary signals exhibit diverse forms in terms of their location, duration, and shape.  
Consequently, simply depicting planetary signals as brief deviation often fails to capture 
their true complexity.

Identifying microlensing planets involves a meticulous procedure that entails considerable 
time and effort. Initially, anomalous events are sought by scrutinizing the light curves 
of lensing events.  Subsequently, the planetary nature of these anomalies is discerned 
through rigorous analyses of the observed light curves. Current lensing surveys annually 
detect more than 3000 events, with roughly 10\% exhibiting anomalies attributed to various 
causes. While some anomalies can be readily attributed to planetary presence, discerning 
the nature of others from their appearance alone is very challenging. To firmly identify 
anomalies, a detailed analysis involving complex procedures and extensive computations is 
necessary. Morphological studies play a crucial role by categorizing anomalies with similar 
characteristics and investigating the origins of each class. This approach not only aids 
in accurately characterizing lensing anomalies for future events with similar structures 
but also facilitates the early diagnosis of anomalies before conducting in-depth analyses.

As the count of microlensing planets rises, planets with signals exhibiting similar anomaly 
patterns are grouped and announced collectively.  \citet{Han2017} and \citet{Poleski2017} 
provided illustrative instances of planetary signals emerging via a recurrent channel, as 
demonstrated in their analysis of the microlensing planets OGLE-2016-BLG-0263Lb and 
MOA-2012-BLG-006Lb, respectively.  \citet{Jung2021} presented planetary signals observed in 
the lensing events OGLE-2018-BLG-0567 and OGLE-2018-BLG-0962, for which the planetary signals 
appeared on the sides of the lensing light curves due to the source stars' crossings over caustics 
situated away from the planet hosts. Additionally, \citet{Han2024} introduced three microlensing 
planets --  MOA-2022-BLG-563Lb, KMT-2023-BLG-0469Lb, and KMT-2023-BLG-0735Lb -- whose signals 
exhibit consistent short-term dip features surrounded by weak bumps on both sides of the dip. 
\citet{Han2023} and \citet{Han2021a} exemplified weak short-term planetary signals generated 
without caustic crossings for the events KMT-2022-BLG-0475, KMT-2022-BLG-1480, KMT-2018-BLG-1976, 
KMT-2018-BLG-1996, and OGLE-2019-BLG-0954.

In this study, we present analyses of four planetary lensing events: KMT-2020-BLG-0757,
KMT-2022-BLG-0732, KMT-2022-BLG-1787, and KMT-2022-BLG-1852. These events exhibit planetary 
signals with a common characteristic, originating from the source crossing over the "planetary" 
caustic induced by "close" planets, followed by source passage through the region of the
"minor-image" perturbation situated between the pair of planetary caustics.  We elucidate 
the technical terms "close," "minor-image," "planetary," and "central" caustics in the 
subsequent section.

The analyses of the planetary events are presented according to the following organization.  
In Sect.~\ref{sec:two}, we provide an overview of the observations conducted for the lensing 
events, including the instrumentation utilized for observations as well as the procedures 
implemented for data reduction and adjustment of error bars. In Sect.~\ref{sec:three}, we 
provide a brief overview on the fundamental principles of planetary microlensing and discuss 
the modeling process used to analyze the observed light curves of the lensing events. Subsequent 
subsections offer comprehensive analyses of individual events and their results. In 
Sect.~\ref{sec:four}, we examine the source stars associated with the events and estimate the 
angular Einstein radii of the events. In Sect.~\ref{sec:five}, we depict the Bayesian analyses 
conducted for each event and present estimates for the physical parameters of the planetary 
systems derived from these analyses. Finally, in Sect.~\ref{sec:six}, we summarize our findings 
and draw conclusions based on the results obtained.

\section{Observations and data}\label{sec:two}

All analyzed lensing events in this work were discovered from the microlensing survey 
conducted toward the Galactic bulge field by the Korea Microlensing Telescope Network 
\citep[KMTNet:][]{Kim2016}.  In Table~\ref{table:one}, we present the equatorial and Galactic 
coordinates of the events, along with their baseline magnitudes ($I_{\rm base}$) and the 
$I$-band extinction ($A_I$) toward the fields. We investigated the availability of additional 
data from other lensing surveys and found that KMT-2020-BLG-0757 was also observed by the 
Microlensing Observations in Astrophysics \citep[MOA:][]{Bond2001} group, who designated the 
event as MOA-2020-BLG-249. For the analysis of this event, we utilized the combined data from 
both the KMTNet and MOA surveys.  After analyzing the KMT-2020-BLG-0757 event, it was discovered 
that additional data had been obtained by the OMEGA group, who conducted follow-up of microlensing 
events in the entire sky.  Our analysis incorporates these newly acquired data.

The KMTNet group utilizes a network of three telescopes that are strategically distributed in 
three locations of the Southern Hemisphere: at the Cerro Tololo Inter-American Observatory in 
Chile (KMTC), the South African Astronomical Observatory in South Africa (KMTS), and the Siding
Spring Observatory in Australia (KMTA). These telescopes have identical specifications, featuring 
a 1.6 m aperture and are each equipped with a camera providing a field of view of 4 square
degrees. The MOA group conducts its survey using the 1.8 m telescope lying at the Mt. John
Observatory located in New Zealand. The MOA telescope is mounted by a camera providing 2.2
square degrees of field of view. Observations by the KMTNet and MOA surveys were mainly done
in the $I$ and the customized MOA-$R$ bands, respectively. For both surveys, a fraction of
images were taken in the $V$ band for the source color measurement. Observational cadences
vary depending on the events and we will mention the cadence when we detail the analysis of
each event.  The supplementary observations of KMT-2020-BLG-0757 by the OMEGA group were 
conducted using the Las Cumbres Observatory Global Telescope (LCOGT) Network, which comprises 
multiple 1-meter telescopes \citep{Brown2013}.

Image reduction and photometry for the lensing events employed automated pipelines tailored 
to each survey. These pipelines utilized code developed by \citet{Albrow2017} for the KMTNet 
survey and by \citet{Bond2001} for the MOA survey. To ensure optimal data usage in the analysis, 
we performed an additional reduction of the KMTNet data employing the photometry code developed 
by \citet{Yang2024}.  The OMEGA data were processed using LCO's BANZAI pipeline \citep{Mccully2018}.  
Following the refinement of the data sets for each event, we adjusted the data error bars.  This 
adjustment aimed not only to maintain consistency with the data scatter but also to ensure that 
the $\chi^2$ value per degree of freedom (dof) was set to unity for each dataset. The normalization 
process followed the procedure described in \citet{Yee2012}.

\section{Analyses of anomalies}\label{sec:three}

Caustics induced by planets are classified into two types based on whether the planet is located
within or outside the Einstein ring of the host star \citep{Dominik1999}.  We use the notation $s$ 
to denote the normalized projected separation between the planet and its host, measured in units of 
the Einstein radius. Both close ($s<1$) and wide ($s>1$) planets generate two sets of caustics: one 
set near the host star (central caustic) and the other set located away from the host at approximately 
${\bf s}-1/{\bf s}$ (planetary caustic). In cases where the planet-to-host mass ratio $q$ is very 
small and the planetary separation $s$ deviates from unity, the central caustics induced by close and 
wide planets closely resemble each other both in shape and size \citep{Griest1998}. Conversely, the 
planetary caustics induced by these two types of planets differ from each other in various aspects: 
a wide planet generates a single four-cusp caustic on the planet side, while a close planet produces 
two sets of three-cusp caustics on the opposite side of the planet. As a result, the planetary signals 
arising from the planetary caustics of these two lens populations exhibit characteristics that can 
nearly always be distinguished.  For an in-depth discussion on the properties of central and planetary 
caustics, refer to \citet{Chung2005} and \citet{Han2006}, respectively.

\begin{figure}[t]
\includegraphics[width=\columnwidth]{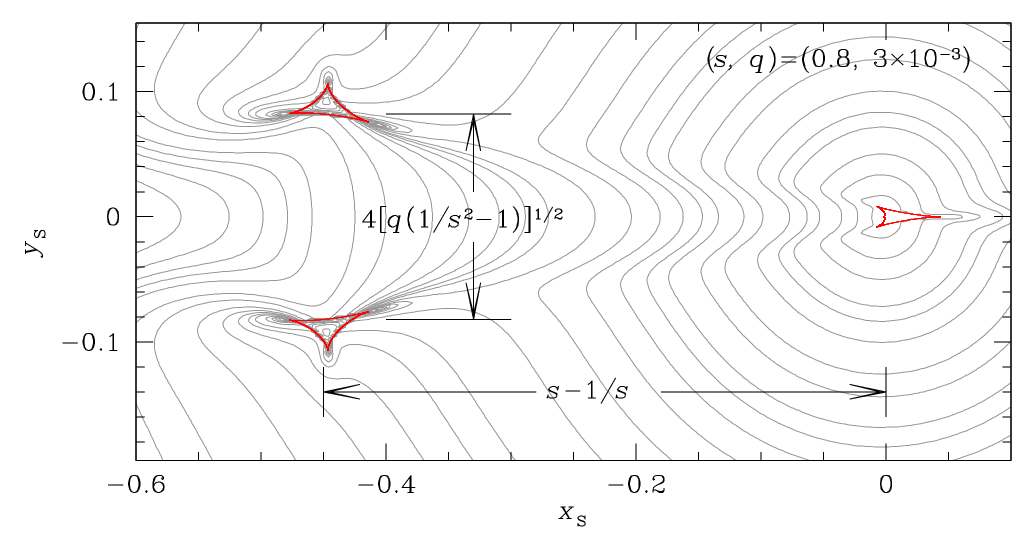}
\caption{
Caustics induced by a close planet and resulting magnification pattern. The red cuspy figures
composed of closed curves represent the caustics, and the grey curves surrounding the caustics 
represent  equi-magnification contours. The planet parameters $(s, q)$ are marked in the upper 
right corner. The width, indicated by arrows, represents the separation between the two sets of 
planetary caustics.  Lengths are scaled to the angular Einstein radius of the lens. 
}
\label{fig:one}
\end{figure}

The planets discussed in this study share common characteristics stemming from perturbations of
the minor image caused by their presence. When a source is gravitationally lensed by a single mass,
it produces two images: the brighter one, known as the "major image," appears outside the
Einstein ring, while the fainter image, referred to as the "minor image," lies within the Einstein
ring \citep{Gaudi1997}. Perturbations to the primary image by a planet result in additional
magnification due to the planet's influence, leading to a positive deviation in the anomaly.
Conversely, perturbations to the minor image by the planet lead to demagnification, resulting in
negative deviations in the anomaly.

In Figure~\ref{fig:one}, we illustrate the magnification pattern of a lens system to show the 
emergence of caustics and the de-magnification region due to the presence of a close 
planet. The planet parameters of the lens system corresponding to the presented configuration are 
$(s, q)=(0.8, 3\times 10^{-3})$, where $q$ denotes the mass ratio between the planet and its host. 
Noteworthy is the expansive span of the de-magnification region, which extends between the central 
and planetary caustics.  This implies that the perturbation resulting from the traversal of a source 
through this area may persist for an extended period.

Assuming a rectilinear relative motion between the lens and the source, a planet-induced anomaly 
in a lensing light curve is characterized by seven fundamental lensing parameters.  Among these, 
the first three parameters $(t_0, u_0, \te)$ describe the approach of the lens and the source. 
Specifically, they represent the time of the closest lens-source approach, the separation (scaled 
to the angular Einstein radius $\thetae$) at $t_0$, and the event timescale, respectively.  The 
subsequent two parameters $(s, q)$ define the planetary lens system, indicating the normalized 
separation and mass ratio between the planet and host.  The next parameter $\alpha$ represents the 
incidence angle of the source relative to the planet-host axis.  As the source crosses the caustic 
induced by the planet, the lensing magnifications are affected by finite-source effects. To incorporate 
these effects, an additional parameter, $\rho$, is necessary. This parameter is defined as the angular 
source radius normalized to $\thetae$ (normalized source radius), that is, $\rho =\theta_*/\thetae$.

For each lensing event, we undertake an analysis to determine a lensing solution, which comprises
the set of lensing parameters that best characterize the observed anomaly. Initially, we search 
for the planet parameters $(s, q)$ using a grid approach with multiple initial values of $\alpha$, 
followed by finding the remaining parameters using a downhill approach based on the Markov chain 
Monte Carlo (MCMC) method. Subsequently, we construct a $\Delta\chi^2$ map on the parameter plane 
of $\log s$--$\log q$ to identify local solutions.  In the subsequent stage, we refine the local 
solutions, and then establish a global solution by comparing the $\chi^2$ values of the local 
solutions.  If the fits of the local solutions are comparable, we present all degenerate solutions 
and investigate the origin of the degeneracy. In the subsequent subsections, we provide detailed 
descriptions of the analyses conducted for the individual events.

\begin{figure}[t]
\includegraphics[width=\columnwidth]{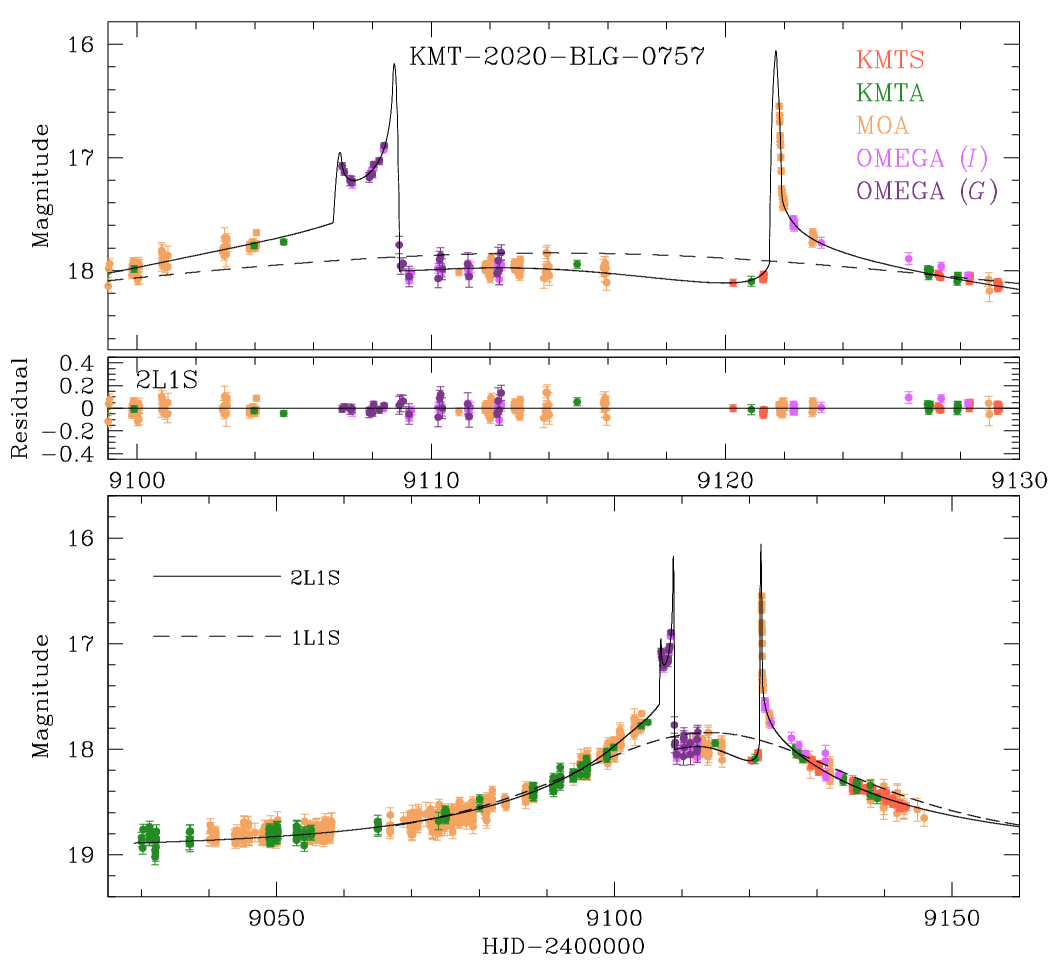}
\caption{
Light curve and models of KMT-2020-BLG-0757. The lower panel provides an overall view, while the 
top panel offers a zoomed-in perspective of the peak region. The dotted and solid lines represent 
models for the inner and outer planetary solutions, respectively.  Below the top panel, two 
additional panels display residuals from these solutions. The dashed curve corresponds to the 
single-lens single-source (1L1S) model.	
}
\label{fig:two}
\end{figure}

\begin{figure}[t]
\includegraphics[width=\columnwidth]{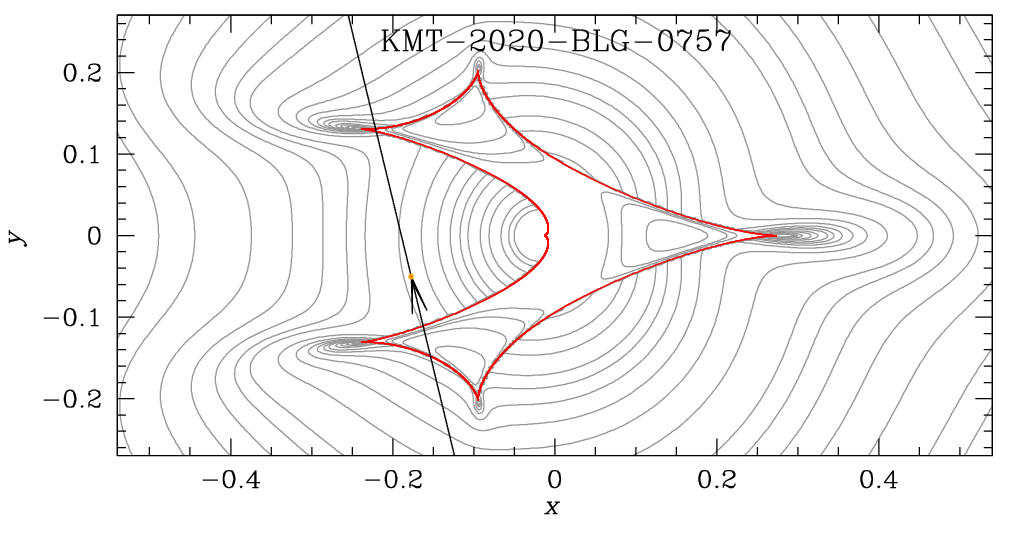}
\caption{
Lens-system configurations of the inner (upper panel) and outer (lower panel) solutions of 
KMT-2020-BLG-0757. In each panel, the red cuspy figure composed of concave curves represent 
caustics and the arrowed line represents the trajectory of the source. Grey curves surrounding 
the caustic represent equi-magnification contours.	
}
\label{fig:three}
\end{figure}

\subsection{KMT-2020-BLG-0757}\label{sec:three-one}

The lensing event KMT-2020-BLG-0757 was first discovered by the KMTNet group on August 28,
2020, corresponding to the reduced heliocentric Julian date ${\rm HJD}^\prime \equiv  {\rm HJD} 
- 2450000 = 9089$.  Subsequently, the MOA group identified the event on September 7, with 
${\rm HJD}^\prime  = 9100$, and designated it as MOA-2020-BLG-249. Following the identification 
reference of the KMTNet survey, which initially discovered the event, we subsequently refer to 
the event as KMT-2020-BLG-0757.  The OMEGA collaboration conducted follow-up observations of 
the event with the alert issued by the MOA group.

\begin{table}[h]
\caption{Lensing solutions of KMT-2020-BLG-0757\label{table:two}}
\begin{tabular*}{\columnwidth}{@{\extracolsep{\fill}}lllcc}
\hline\hline
\multicolumn{1}{c}{Parameter}    &
\multicolumn{1}{c}{Value}        \\
\hline
  $\chi^2$               &   $1662.9             $  \\   
  $t_0$ (HJD$^\prime$)   &   $9112.475 \pm 0.066 $  \\   
  $u_0$                  &   $0.1863 \pm 0.0011  $  \\    
  $\te$ (days)           &   $51.88 \pm 1.15     $  \\      
  $s$                    &   $0.9463 \pm 0.0043  $  \\       
  $q$ (10$^{-3}$)        &   $17.06 \pm 0.33     $  \\     
  $\alpha$ (rad)         &   $1.3374 \pm 0.0073  $  \\   
  $\rho$ (10$^{-3}$)     &   $2.12 \pm 0.11      $  \\      
\hline                                                   
\end{tabular*}
\tablefoot{ ${\rm HJD}^\prime = {\rm HJD}- 2450000$.  }
\end{table}

Figure~\ref{fig:two} illustrates the event light curve constructed from the combination of the 
KMTNet, MOA, and OMEGA data.  We note that the light-curve coverage of events from the 2020 
season (including KMT-2020-BLG-0757) was severely impacted by the Covid-19 pandemic. Shortly after 
the start of the season, observations were discontinued at KMTC and KMTS. They were only resumed 
for KMTS near the end of the season, at HJD$^\prime$ $\sim 9120$, as can be seen in Figure~\ref{fig:two}.  
As a result, the majority of the light curve was captured by the MOA and KMTA datasets, which 
were operation during the season.  No data are available after ${\rm HJD}^\prime  \sim 9144$ due 
to the conclusion of the bulge season.  The anomalous nature of the event was identified from the 
sharp rises of the source flux around ${\rm HJD}^\prime \sim 9109$ and $\sim 9121$, which were 
observed by the OMEGA and MOA groups, respectively.  Later examination of the light curve revealed 
additional deviations from the single-source single-lens (1L1S) model: positive deviations in the 
range $9100 \lesssim {\rm HJD}^\prime \lesssim 9105$ and negative deviations during $9110 \lesssim 
{\rm HJD}^\prime \lesssim 9121$.  From the investigation of previously identified microlensing planets, 
we found that the anomaly feature was similar to those appeared in the events MOA-2009-BLG-387 
\citep{Batista2011}, OGLE-2015-BLG-0051 \citep{Han2016}, MOA-2016-BLG-227 \citep{Koshimoto2017}, 
KMT-2017-BLG-1038, KMT-2017-BLG-1146 \citep{Shin2019}, KMT-2017-BLG-2509, and OGLE-2019-BLG-0299 
\citep{Han2021b}.

Through the modeling of the light curve, we identified a unique solution with $(s, q) \sim (0.95, 
17.1\times 10^{-3})$.  The complete lensing parameters of the solution, along with the corresponding 
$\chi^2$ value of the fit, are listed in Table~\ref{table:two}.  In Figure~\ref{fig:three}, 
we illustrate the configuration of the lens system corresponding to the solution.  The configuration 
shows that the planetary and central caustics merge to form a single resonant caustic.  The source 
crossed the caustic four times: at ${\rm HJD}^\prime \sim 9106.5$, $\sim 9108.5$, $\sim 9121.3$, and 
$\sim 9122.0$.  The first and second passages occurred as the source traversed into and out of the 
lower planetary caustic, while the third and fourth passages occurred as it traversed into and out 
of upper planetary caustic.  The U-shape trough region of the first caustic-crossing pair was covered 
by the OMEGA data, and the caustic exit of the second caustic-crossing pair was resolved by the MOA 
data.  The normalized source radius was determined through the resolved caustic during the last 
crossing.  The extended negative deviation region observed during the period $9112\lesssim  
{\rm HJD}^\prime \lesssim  9121$ was attributed to perturbations of the minor image resulting from 
the source passing through the region between the upper and lower planetary caustics.

\begin{figure}[t]
\includegraphics[width=\columnwidth]{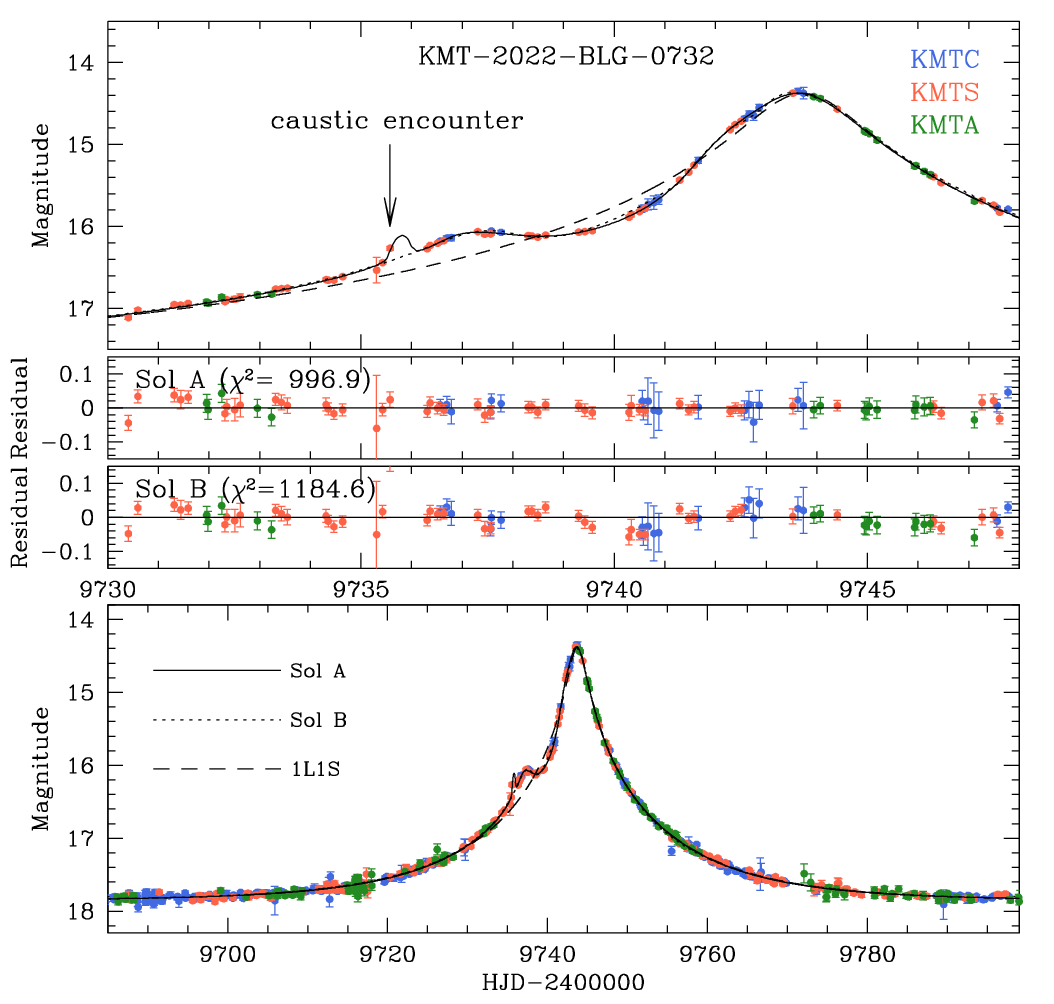}
\caption{
Light curve of the lensing event KMT-2022-BLG-0732. The notation used here is consistent with that 
employed in Fig.~\ref{fig:two}.  The arrow in the upper panel  marks the KMTS data corresponding 
to the caustic encounter at ${\rm HJD}^\prime = 9735.575$.
}
\label{fig:four}
\end{figure}

\subsection{KMT-2022-BLG-0732}\label{sec:three-two}

The KMTNet team detected the source flux enhancement of KMT-2022-BLG-0732 caused by lensing during 
the early stage of the event on May 9, 2022, corresponding to ${\rm HJD}^\prime = 9709$. The source 
was located within the KMTNet BLG37 field, toward which observations were conducted at a cadence of 
2.5 hours.  This event was not reported by other survey initiatives.  Figure~\ref{fig:four} presents 
the light curve of KMT-2022-BLG-0732. The rising portion of the light curve showcases a complex 
anomaly pattern.  Especially, the segment observed during the period $9733.8 \lesssim {\rm HJD}^\prime 
\lesssim 9738.5$ exhibits positive deviations with respect to a 1L1S model, while the segment observed 
during the period $9738.5 \lesssim {\rm HJD}^\prime \lesssim 9741.0$ displays negative deviations.

By analyzing the lensing light curve, we have identified two distinct local solutions characterized 
by the planetary parameters $(s, q)_{\rm A}\sim (0.81, 2.01\times 10^{-3})$ and $(s, q)_{\rm B}\sim 
(1.02, 2.01\times 10^{-3})$.  These solutions are respectively referred to as "sol A" and "sol B." 
Both solutions exhibit very low mass ratios between the lens components, indicating that the companion 
to the lens is likely a planetary mass object. Detailed lensing parameters of the solutions are provided 
in Table~\ref{table:three}, while the model curves and residuals are depicted in Figure~\ref{fig:four}.  
Upon comparing the fits, it is evident that sol A offers a better explanation for the observed anomaly, 
particularly during the negative deviation phase $(9738.5 \lesssim {\rm HJD}^\prime \lesssim 9741.0)$.  
This is further supported by the substantial difference in $\chi^2$ values ($\Delta\chi^2 = 186.7$) 
between the two solutions, leading us to conclusively reject sol B.

\begin{figure}[t]
\includegraphics[width=\columnwidth]{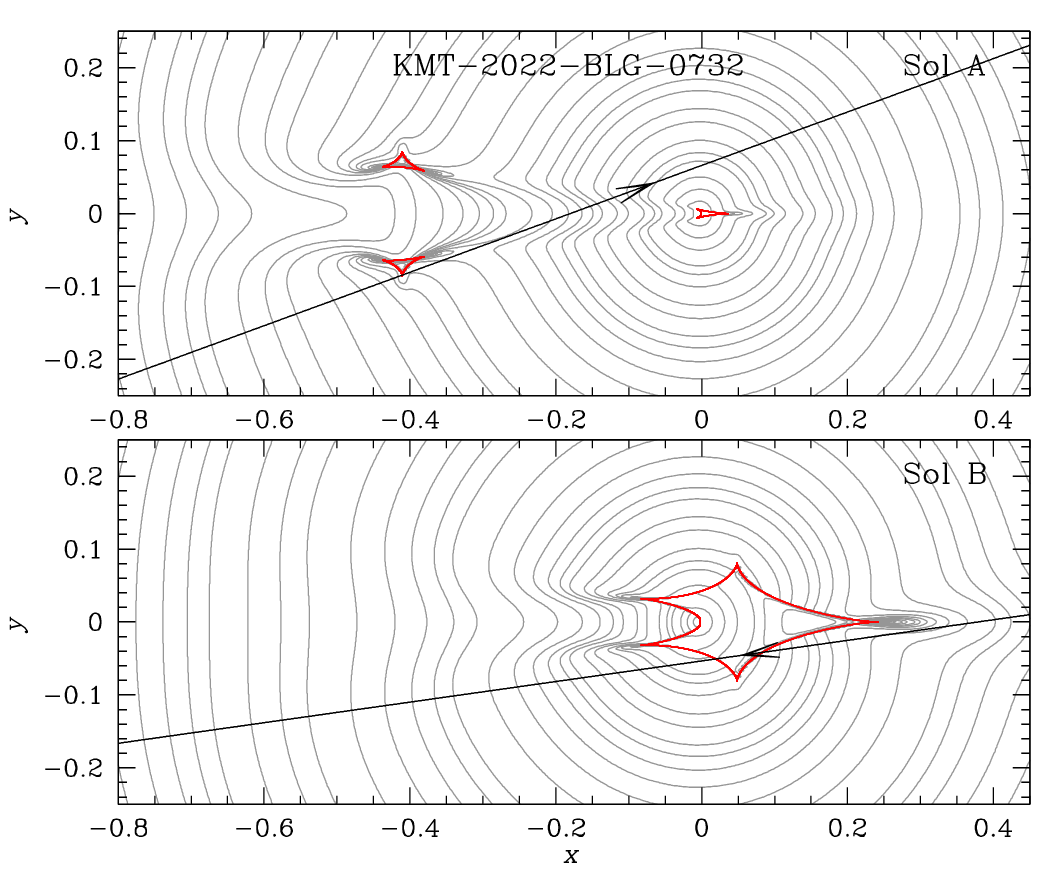}
\caption{
Configuration of lens system corresponding to sol A and sol B of KMT-2022-BLG-0732. Notations are 
same as those in Fig.~\ref{fig:three}.	
}
\label{fig:five}
\end{figure}

\begin{table}[t]
\caption{Lensing parameters of KMT-2022-BLG-0732\label{table:three}}
\begin{tabular*}{\columnwidth}{@{\extracolsep{\fill}}lllcc}
\hline\hline
\multicolumn{1}{c}{Parameter}    &
\multicolumn{1}{c}{sol A}        &
\multicolumn{1}{c}{sol B}        \\
\hline
  $\chi^2$               &  $996.9               $ &  $1183.6              $   \\   
  $t_0$ (HJD$^\prime$)   &  $9743.6907 \pm 0.0089$ &  $9743.6507 \pm 0.0096$   \\   
  $u_0$                  &  $0.0619 \pm 0.0011   $ &  $0.0531 \pm 0.0009   $   \\   
  $\te$ (days)           &  $19.18 \pm 0.18      $ &  $21.09 \pm 0.20      $   \\   
  $s$                    &  $0.8148 \pm 0.0016   $ &  $1.0233 \pm 0.0015   $   \\   
  $q$ (10$^{-3}$)        &  $2.007 \pm 0.067     $ &  $2.009 \pm 0.066     $   \\   
  $\alpha$ (rad)         &  $2.7905 \pm 0.0044   $ &  $3.5641 \pm 0.0017   $   \\   
  $\rho$ (10$^{-3}$)     &  $11.35 \pm 1.23      $ &  $11.51 \pm 0.75      $   \\   
\hline                                                   
\end{tabular*}
\end{table}

The lens system configuration for the event is depicted in Figure~\ref{fig:five}. Although ruled out, 
we also present the configuration corresponding to sol B to elucidate the origin of the degeneracy 
between the two solutions.  The interpretation of sol~A bears resemblance to that of KMT-2020-BLG-0757, 
as it involves the source passing through one of the planetary caustics induced by a close planet, 
followed by the passage through the minor-image perturbation region formed between the planetary and 
central perturbation region.  We mark the KMTS data corresponding to the caustic encounter at 
${\rm HJD}^\prime = 9735.575$ with an arrow. Since the caustic was covered by a single point, we 
verified its validity by performing an additional modeling without this point. The best-fit model 
remained virtually unchanged, confirming the authenticity of the data point.  According to sol B, 
on the other hand, the positive deviation is attributed to the source passage through the region 
extending from the on-line cusp of a resonant caustic, while the negative deviation is explained 
by the source passage near an off-axis cusp of the caustic. The resemblance between the model 
curves of sol A and sol B is not attributable to inherent degeneracy within the system, nor is 
the degeneracy notably significant. Therefore, we can classify it as an accidental degeneracy. A 
notable distinction in the anomaly pattern of KMT-2022-BLG-0732 compared to that of KMT-2020-BLG-0757 
is the absence of an evident caustic-crossing spike.  Upon examining the anomaly region corresponding 
to the time of the caustic crossing, we found that the caustic spike was blurred out because of 
significant finite-source effects. From the analyses of the region of the anomaly that was impacted 
by finite-source effects, the normalized source radius, $\rho = (11.35 \pm 1.23)\times 10^{-3}$, was 
measured.

\begin{figure}[t]
\includegraphics[width=\columnwidth]{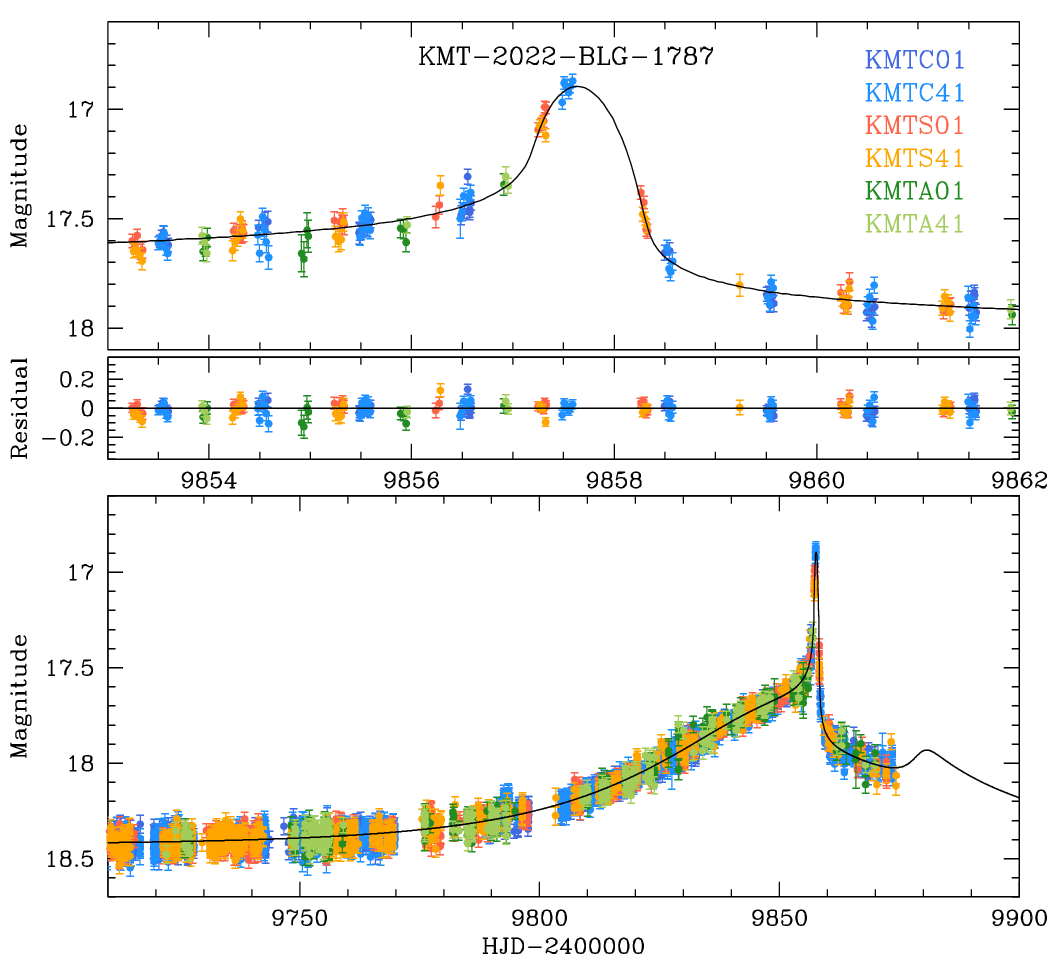}
\caption{
Lensing light curve KMT-2022-BLG-1787.  The notation used here is consistent with that
employed in Fig.~\ref{fig:two}.	
}
\label{fig:six}
\end{figure}

\subsection{KMT-2022-BLG-1787}\label{sec:three-three}

The lensing event KMT-2022-BLG-1787 was initially detected by the KMTNet group on August 16, 2022, 
corresponding to ${\rm HJD}^\prime  = 9808$. The event was densely observed because the source was 
located within the overlapping region of KMTNet's prime fields BLG01 and BLG41, toward which 
observations were conducted with a cadence of 0.5 hours for each individual field and 0.25 hours 
in combined mode. Figure~\ref{fig:six} shows the lensing light curve of the event.  Continuing until 
the end of the 2022 bulge season, observations of its later stages were limited. The light curve 
exhibits anomalies similar to those observed in the two previous events, featuring both positive 
(around ${\rm HJD}^\prime  \sim  9857.5$) and negative deviations (occurring after ${\rm HJD}^\prime 
\sim 9859$). This suggests a potential origin of the anomaly akin to those observed in the prior events.

\begin{figure}[t]
\includegraphics[width=\columnwidth]{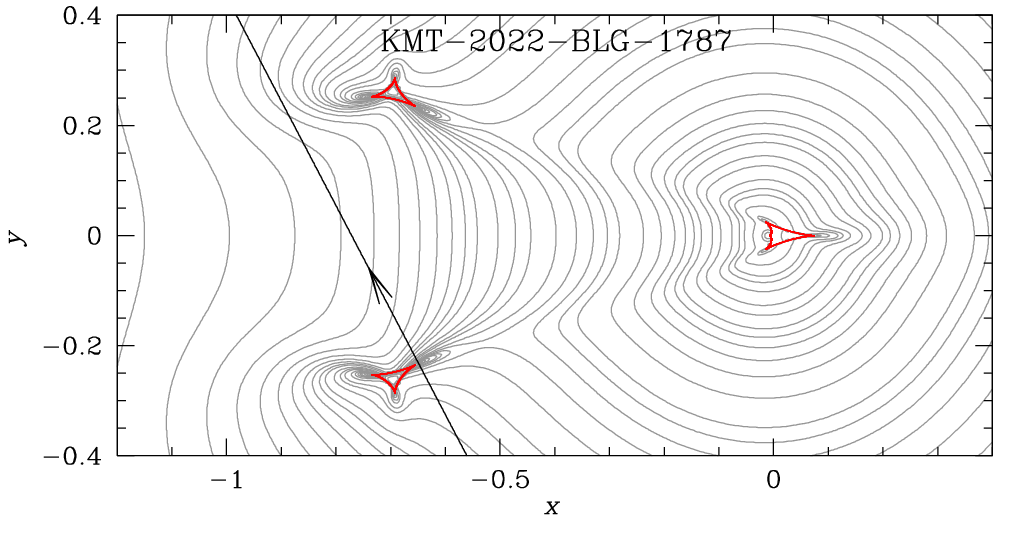}
\caption{
Lens-system configuration of KMT-2022-BLG-1787. Notations are same as those in Fig.~\ref{fig:two}.
}
\label{fig:seven}
\end{figure}

\begin{table}[t]
\caption{Lensing parameters of KMT-2022-BLG-1787\label{table:four}}
\begin{tabular*}{\columnwidth}{@{\extracolsep{\fill}}lllcc}
\hline\hline
\multicolumn{1}{c}{Parameter}    &
\multicolumn{1}{c}{Value}        \\
\hline
  $\chi^2$               &  $5448.1           $   \\   
  $t_0$ (HJD$^\prime$)   &  $9853.75 \pm 0.19 $   \\   
  $u_0$                  &  $0.682 \pm 0.020  $   \\   
  $\te$ (days)           &  $41.12 \pm 0.91   $   \\   
  $s$                    &  $0.7037 \pm 0.0066$   \\   
  $q$ (10$^{-3}$)        &  $15.07 \pm 0.86   $   \\   
  $\alpha$ (rad)         &  $1.056 \pm 0.013  $   \\   
  $\rho$ (10$^{-3}$)     &  $14.07 \pm 0.40   $   \\   
\hline                                                   
\end{tabular*}
\end{table}

In line with the anticipated anomaly pattern, the modeling of the event's light curve yielded results
that are consistent with those observed in the previous events. This reaffirms that the anomaly was
caused by the source crossing over a planetary caustic generated by a close planet, followed by its
passage through the region affected by the minor-image perturbation.  Analysis of the event yielded 
a unique solution without any degeneracy.  The estimated planet parameters are $(s, q) \sim (0.70, 
15.1\times 10^{-3})$. The complete list of the lensing parameters is provided in Table~\ref{table:four}.

In Figure~\ref{fig:seven}, we present the lens-system configuration corresponding to the solution. 
The caustic configuration is very similar to that of KMT-2022-BLG-0732, except that the width between 
the planetary caustics is wider due to the larger planet-to-host mass ratio. The source passage through 
the tip of the lower planetary caustic produced the positive peak at ${\rm HJD}^\prime \sim 9857.5$, 
and the subsequent passage through the minor-image perturbation region resulted in a negative deviation. 
From analyzing the distorted bump feature due to finite-source effects, the normalized source radius 
was securely measured to be $\rho = (14.07 \pm 0.40)\times 10^{-3}$. It is worth noting that had 
observations continued until the end of the event, a second weak bump would have been visible around 
${\rm HJD}^\prime = 9881$.

\begin{figure}[t]
\includegraphics[width=\columnwidth]{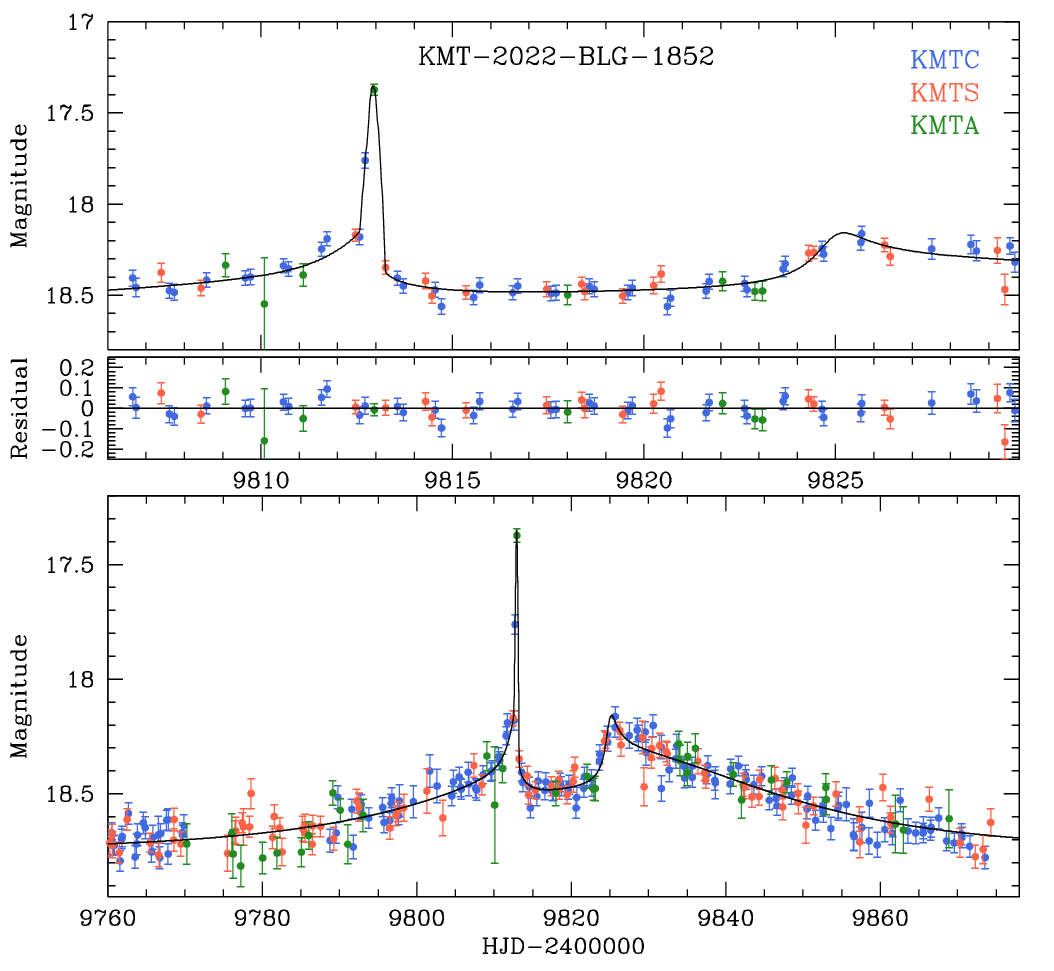}
\caption{
Light curve of KMT-2022-BLG-1852. The notation used here is consistent with that employed in 
Fig.~\ref{fig:two}.
}
\label{fig:eight}
\end{figure}

\subsection{KMT-2022-BLG-1852}\label{sec:three-four}

The KMTNet group detected the lensing event KMT-2022-BLG-1852 on August 19, 2022 (${\rm HJD}^\prime = 
9811$). The source was positioned within the KMTNet peripheral field BLG36, toward which observations 
were conducted with the lowest cadence (5.0 hours) among all 27 KMTNet fields.  Despite this limitation 
in cadence, the light curve depicted in Figure~\ref{fig:eight} clearly exhibits an anomaly.  This anomaly 
consists of an elongated dip ($9733.8 \lesssim {\rm HJD}^\prime \lesssim 9738.5$) flanked by bumps on 
either side, resembling that observed in KMT-2022-BLG-1787. The significant deviation observed in the 
pre-dip bump suggests its formation due to a caustic crossing, while the weaker post-dip bump is likely 
a result of the source approaching a caustic cusp.  These features in the anomaly is similar to those 
of KMT-2022-BLG-1787.

As expected due to the similarity in anomaly patterns, modeling the light curve of KMT-2022-BLG-1852 
produced results consistent with those obtained for KMT-2022-BLG-1787: indicating the presence of a 
close planet within the lens system, and the anomaly being caused by the source traversing the planetary 
caustic followed by the passage through the region of minor-image perturbations. The modeling resulted 
in a single unique solution with planet parameters $(s, q) \sim (0.67, 8.7\times 10^{-3})$. A detailed 
list of the lensing parameters is provided in Table~\ref{table:five}.

\begin{table}[t]
\caption{Lensing parameters of KMT-2022-BLG-1852\label{table:five}}
\begin{tabular*}{\columnwidth}{@{\extracolsep{\fill}}lllcc}
\hline\hline
\multicolumn{1}{c}{Parameter}    &
\multicolumn{1}{c}{Value}        \\
\hline
  $\chi^2$               &  $463.7            $    \\   
  $t_0$ (HJD$^\prime$)   &  $9825.55 \pm 0.36 $    \\   
  $u_0$                  &  $0.746 \pm 0.048  $    \\   
  $\te$ (days)           &  $30.06 \pm 1.22   $    \\   
  $s$                    &  $0.666 \pm 0.015  $    \\   
  $q$ (10$^{-3}$)        &  $8.74 \pm 0.49    $    \\   
  $\alpha$ (rad)         &  $1.833 \pm 0.017  $    \\   
  $\rho$ (10$^{-3}$)     &  $6.62 \pm 1.29    $    \\   
\hline                                                   
\end{tabular*}
\end{table}

\begin{figure}[t]
\includegraphics[width=\columnwidth]{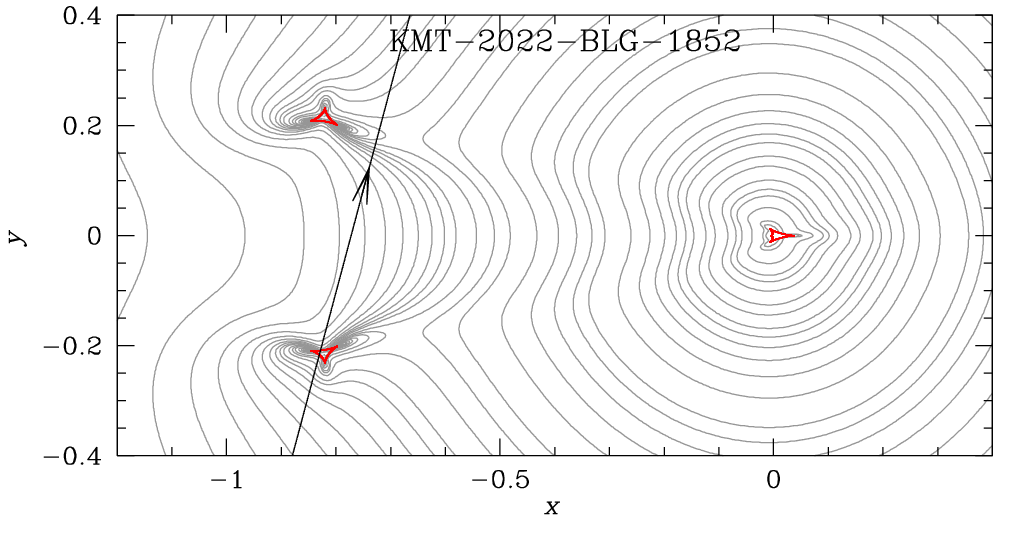}
\caption{
Lens-system configuration of KMT-2022-BLG-1852. 
}
\label{fig:nine}
\end{figure}

Figure~\ref{fig:nine} illustrates the configuration of the lens system.  The configuration closely 
resembles that of KMT-2022-BLG-1787.  One minor distinction is the direction of the source trajectory: 
in KMT-2022-BLG-1852, it is directed toward the right, while in KMT-2022-BLG-1787, it is directed 
toward the left.  Despite the relatively low cadence, the first bump, which occurred as a result 
of the source crossing over the planetary caustic, was captured by the combined data sets. Consequently, 
the normalized source radius was measured to be $\rho  = (6.62 \pm 1.29) \times 10^{-3}$, albeit with
a relatively large uncertainty.  As a somewhat technical point, the careful reader may have noticed 
that the two KMTC points on the night of HJD'=9812.xx (i.e., the caustic rise) are separated by just 
3.4 hours, despite the nominal cadence of 5 hours for BLG36. Indeed such "anomalously short" observation 
intervals are present in all neighboring nights. The reason for this is that during the second half of 
the season, KMTNet runs $\sim 20$ minute cycle of observations of its Eastern fields at the end of the 
night, that is, when the full set of KMT fields can no longer be observed because the Western fields 
have set. For typical fields with cadences of 1 hour or 2.5 hours, this results in a modest increase 
in coverage, but the effect is more dramatic for BLG36 because it has a 5-hour cadence. In this
case, this end-of-night "extra" point became one of only two points on the caustic, whose inclusion
thus allowed for a good measurement of $\rho$.

\begin{table*}[t]
\caption{Source parameters, angular source radii, Einstein radii, and relative lens-source proper motions\label{table:six}}
\begin{tabular}{lllll}
\hline\hline
\multicolumn{1}{c}{Quantity}            &
\multicolumn{1}{c}{KMT-2020-BLG-0757}   &
\multicolumn{1}{c}{KMT-2022-BLG-0732}   &
\multicolumn{1}{c}{KMT-2022-BLG-1787}   &
\multicolumn{1}{c}{KMT-2022-BLG-1852}   \\
\hline
$(V-I)_{\rm S}$             &  $1.506 \pm 0.049 $  &  $2.761 \pm 0.045 $  &  $3.515 \pm 0.098 $  &  $1.443 \pm 0.261 $  \\
$I_{\rm S}$                 &  $19.322 \pm 0.007$  &  $17.467 \pm 0.002$  &  $18.423 \pm 0.006$  &  $19.560 \pm0.032 $  \\
$(V-I, I)_{\rm RGC}$        &  $(1.948, 15.704) $  &  $(2.733, 17.091) $  &  $(3.512, 17.385) $  &  $(1.532, 16.569) $  \\
$(V-I, I)_{{\rm RGC},0}$    &  $(1.060, 14.588) $  &  $(1.060, 14.619) $  &  $(1.060, 14.486) $  &  $(1.060, 14.218) $  \\
$(V-I)_{{\rm S},0}$         &  $0.619 \pm 0.063 $  &  $1.087 \pm 0.060 $  &  $1.063 \pm 0.106 $  &  $0.971 \pm 0.264 $  \\
$I_{{\rm S},0}$             &  $18.206 \pm 0.021$  &  $14.995 \pm 0.020$  &  $15.524 \pm 0.021$  &  $17.210 \pm 0.038$  \\
Source type                 &   F9V                &   K3.5III            &   K3.5III            &   K2.5V (subgiant)   \\
$\theta_*$ ($\mu$as)        &  $0.651 \pm 0.061 $  &  $4.901 \pm 0.452 $  &  $3.75 \pm 0.48   $  &  $1.53 \pm 0.42   $  \\
$\thetae$ (mas)             &  $0.381 \pm 0.045 $  &  $0.433 \pm 0.063 $  &  $0.266 \pm 0.036 $  &  $0.231 \pm 0.078 $  \\
$\mu$ (mas/yr)              &  $2.35 \pm 0.28   $  &  $8.25 \pm 1.20   $  &  $2.37 \pm 0.31   $  &  $2.81 \pm 0.94   $  \\
\hline
\end{tabular}
\end{table*}

\section{Source stars and angular Einstein radii}\label{sec:four}

In this section, we provide details regarding the source stars involved in the individual lensing
events. With the specified source type, we estimated the angular Einstein radius and relative
lens-proper motion using the relations:
\begin{equation}
\thetae = {\theta_* \over \rho};\qquad
\mu = {\thetae \over \te},
\label{eq1}
\end{equation}
where $\theta_*$ represents the angular radius of the source. The normalized source radius $\rho$ was 
derived by examining the section of the light curve corresponding to caustic crossings, whereas the 
angular source radius $\theta_*$ was inferred from the type of the source. Given that all events 
exhibited anomalies involving caustic crossings, we were able to constrain the normalized source 
radii and subsequently the Einstein radii.

\begin{figure}[t]
\includegraphics[width=\columnwidth]{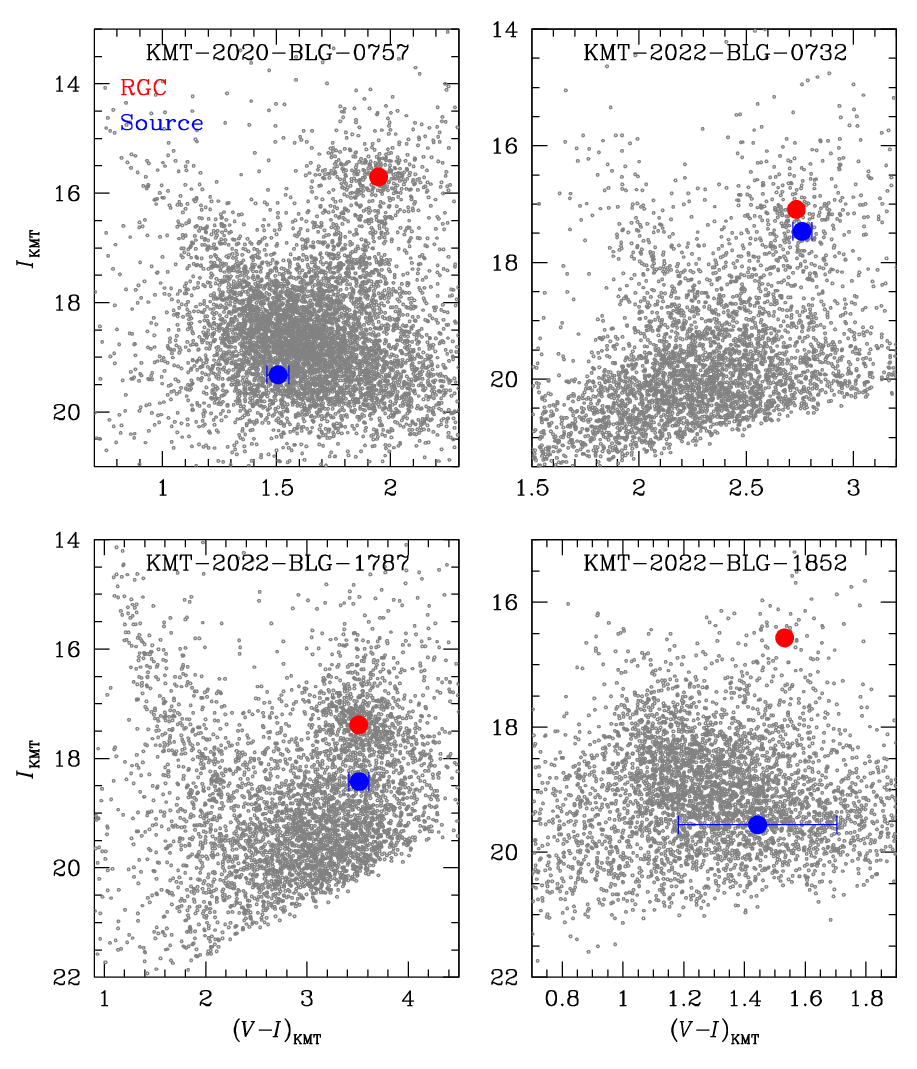}
\caption{
Locations of the source and red giant clump (RGC) centroid in the instrumental
color-magnitude diagrams of the lensing events KMT-2020-BLG-0757, KMT-2022-BLG-0732,
KMT-2022-BLG-1787, and KMT-2022-BLG-1852. 
}
\label{fig:ten}
\end{figure}

We determined the source type employing the methodology outlined in \citet{Yoo2004}.  According 
to this method, we first measured the instrumental color and magnitude, $(V-I, I)_{\rm S}$, of 
the source, and placed the source in the color-magnitude diagram (CMD) of stars lying near the 
source.  The $I$- and $V$-band source magnitudes were determined by fitting the light curves of 
the corresponding bands to the model. This process utilized the data processed with the pyDIA 
code \citep{Albrow2017}.  We then calibrated the color and magnitude using the centroid of 
the red giant clump (RGC) in the CMD as a reference, that is, 
\begin{equation} 
(V-I, I)_{{\rm S},0} = (V-I, I)_{{\rm RGC},0} + \Delta(V-I, I).  
\label{eq2} 
\end{equation} 
Here $(V-I, I)_{{\rm S},0}$ and $(V-I, I)_{{\rm RGC},0}$ respectively represent the reddening 
and extinction-corrected colors and magnitudes of the source and RGC centroid, and the term $\Delta(V-I, I)
= (V-I, I)_{\rm S}-(V-I)_{\rm RGC}$ indicates the offsets in color and magnitude between the source 
and RGC centroid. The RGC centroid can be used as a reference for calibration because its de-reddened 
color and magnitude are known from \citet{Bensby2013} and \citet{Nataf2013}, respectively.

Figure~\ref{fig:ten} shows the source stars and RGC centroids in the CMDs for the lensing events. 
Table~\ref{table:six} presents the values of $(V-I, I)_{\rm S}$, $(V-I, I)_{\rm RGC}$, $(V-I, 
I)_{{\rm RGC},0}$, and $(V-I, I)_{{\rm S},0}$ for the events, along with the corresponding spectral 
types of the source stars. The spectral types inferred from the measured colors and magnitudes 
are F9V for KMT-2020-BLG-0757, K3.5III for KMT-2022-BLG-0732 and KMT-2022-BLG-1787, and K2.5 
subgiant for KMT-2022-BLG-1852.

The angular source radius was determined based on the measured source color and magnitude. For
this determination, we initially converted the $V-I$ color to $V-K$ color using the color-color
relation provided by \citet{Bessell1988}. Subsequently, we utilized the $(V-K, I)$--$\theta_*$ 
relationship established by \citet{Kervella2004} to infer the angular source radius. With the 
determined angular source radius, the Einstein radius and proper motion were calculated using 
the relations in Eq.~(\ref{eq1}). We present the estimated values of $\theta_*$, $\thetae$ and 
$\mu$ in Table~\ref{table:six}.  We further checked the information on the source stars in the 
{\it Gaia} catalog \citep{Gaia2018}.  For KMT-2022-BLG-0732 and KMT-2022-BLG-1787, we identified 
the source stars, while those of KMT-2020-BLG-0757 and KMT-2022-BLG-1852 were not registered in 
the catalog. Even for the events with identified source stars, only the $G$-band magnitudes were 
available, with no information specifying the spectral types of the source stars.

\begin{table*}[t]
\caption{Physical lens parameters\label{table:seven}}
\begin{tabular}{lllll}
\hline\hline
\multicolumn{1}{c}{Quantity}            &
\multicolumn{1}{c}{KMT-2020-BLG-0757}   &
\multicolumn{1}{c}{KMT-2022-BLG-0732}   &
\multicolumn{1}{c}{KMT-2022-BLG-1787}   &
\multicolumn{1}{c}{KMT-2022-BLG-1852}   \\
\hline
 $M_{\rm host}$ ($M_\odot$)       & $0.58^{+0.33}_{-0.30}     $   &  $0.54^{+0.31}_{-0.31}$    & $0.42^{+0.32}_{-0.23}$     &  $0.32^{+0.34}_{-0.19}   $ \\ [0.4ex]
 $M_{\rm planet}$ ($M_{\rm J}$)   & $10.71^{+6.17}_{-5.61}    $   &  $1.12^{+0.65}_{-0.65}$    & $6.64^{+4.98}_{-3.64}$     &  $4.98^{+5.42}_{-2.94}   $ \\ [0.4ex]
 $\dl$ (kpc)                      & $6.67^{+0.93}_{-1.30}     $   &  $6.66^{+1.19}_{-1.84}$    & $7.55^{+0.89}_{-1.30}$     &  $6.27^{+0.90}_{-1.15}   $ \\ [0.4ex]
 $a_\perp$ (AU)                   & $11.54^{+1.61}_{-2.25}    $   &  $4.11^{+0.74}_{-1.14}$    & $6.30^{+0.75}_{-1.08}$     &  $5.03^{+0.72}_{-0.92}   $ \\ [0.4ex]
 $p_{\rm disk}$                   & $31\%                     $   &  $64\%                $    & $25\%                $     &  $26\%                   $ \\ [0.4ex]
 $p_{\rm bulge}$                  & $69\%                     $   &  $36\%                $    & $75\%                $     &  $74\%                   $ \\ [0.4ex]
\hline
\end{tabular}
\end{table*}

\section{Physical planet parameters}\label{sec:five}

In this section, we derive estimates for the mass $M$ and distance $\dl$ of the discovered 
planetary systems. These physical lens parameters were derived from the lensing observables, 
namely the event timescale and the angular Einstein radius. These observables can provide constraints
on the physical parameters because they are connected to $M$ and $\dl$ through the relationships
represented as:
\begin{equation}
\te = {\thetae \over \mu};\qquad 
\thetae = (\kappa M \pi_{\rm rel})^{1/2}.
\label{eq3}
\end{equation}
Here $\kappa = 4G/(c^2{\rm AU}) \simeq 8.14~{\rm mas}/M_\odot$, $\pi_{\rm rel} = \pi_{\rm L} 
- \pi_{\rm S} = {\rm AU}(1/\dl - 1/\ds)$ represents the relative
lens-source parallax, and $\ds$ denotes the source distance. The lens mass and distance can be
uniquely determined by measuring an additional lensing observable of the microlens parallax $\pie$
through the relations given by \citet{Gould2000} as:
\begin{equation}
M={ \thetae \over \kappa \pie};\qquad
\dl = {{\rm AU} \over \pie\thetae + \pi_{\rm S} }.
\label{eq4}
\end{equation}
The microlens-parallax vector is defined as $\pivec_{\rm E} = (\pi_{\rm rel}/\thetae)(\muvec/\mu)$, 
and its value can be determined by observing the subtle deformation of a lensing light curve, resulting 
from the deviation of the relative lens-source motion from rectilinear, caused by the orbital motion of 
Earth around the Sun: microlens-parallax effects \citep{Gould1992b}.  Because none of the events had a 
measured microlens parallax, we employed Bayesian analysis to estimate the lens parameters. This analysis 
incorporates the constraints provided by the measured lensing observables $\te$ and $\thetae$ together 
with priors of a Galaxy model and a mass function of objects within the Galaxy.

\begin{figure}[t]
\includegraphics[width=\columnwidth]{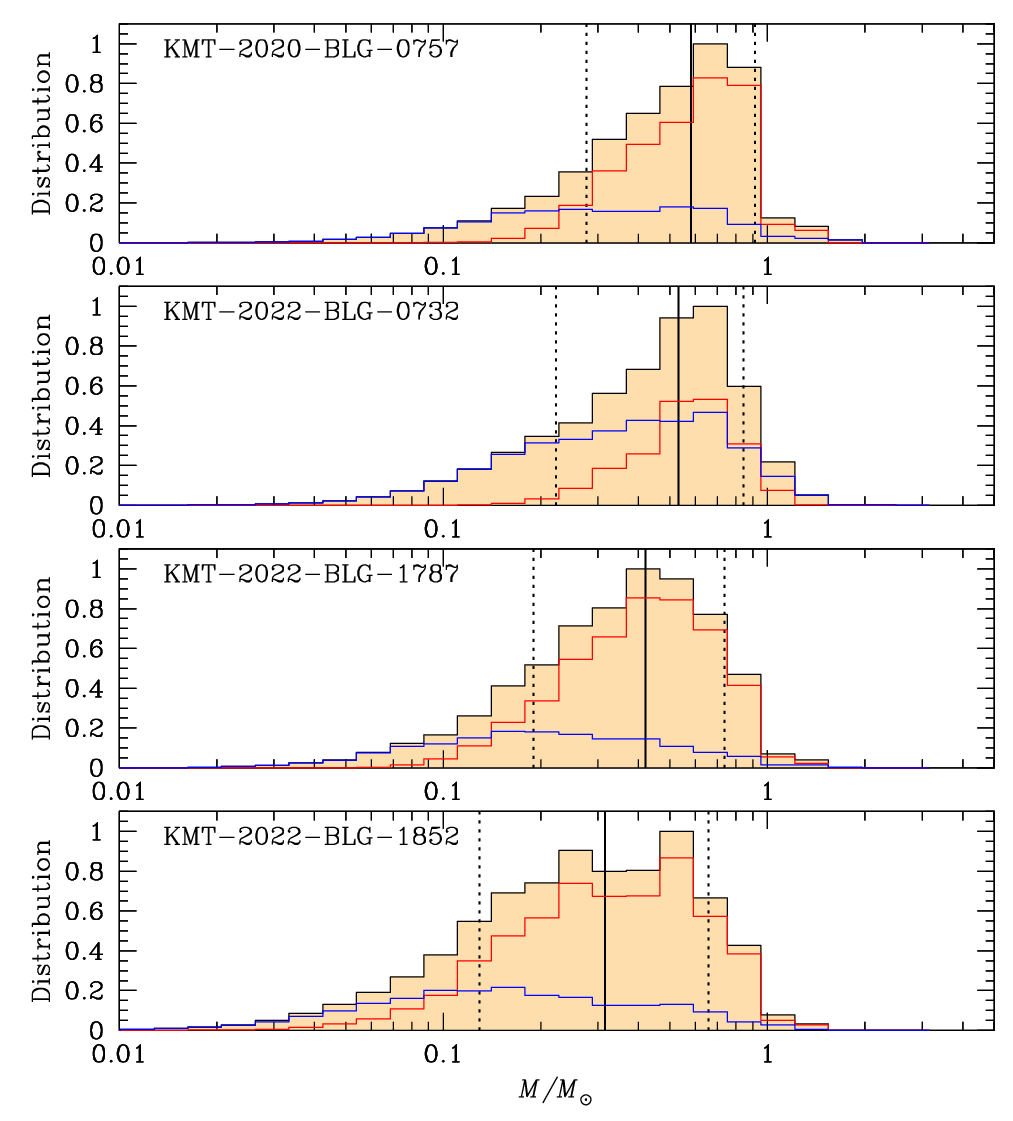}
\caption{
Posteriors for the masses of the planetary systems estimated from Bayesian analyses. Within each 
distribution, the median value is denoted by a solid vertical line, while the uncertainty range 
is depicted by two dotted vertical lines. The contributions from the disk and bulge lens populations 
are respectively shown in blue and red curves, with the combined contribution represented by the 
black curve.}
\label{fig:eleven}
\end{figure}

\begin{figure}[t]
\includegraphics[width=\columnwidth]{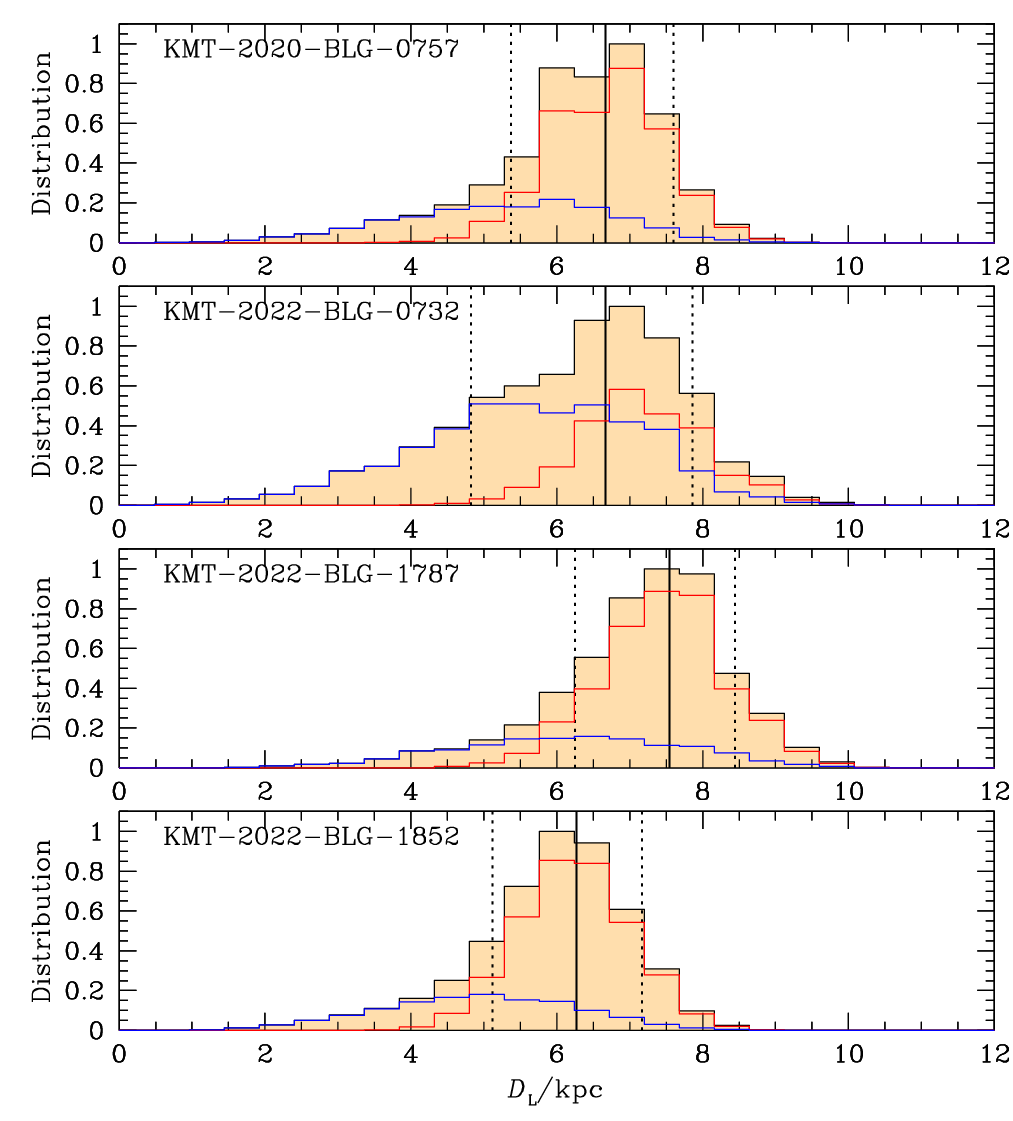}
\caption{
\label{fig:twelve}
Posteriors for the distances to the planetary systems. The notations used are consistent
with those in Fig.~\ref{fig:eleven}.
}
\end{figure}

We carried out the Bayesian analysis according to the following procedure. In the initial stage, 
we conducted a Monte Carlo simulation to generate a large number of synthetic lensing events. Within 
this simulation, we extracted the distances to the lens and source, along with their relative proper 
motion from a Galactic model. Additionally, we derived the lens mass from a mass function model.  
Specifically, we employed the Galactic model outlined in \citet{Jung2021} to derive $\dl$, $\ds$, and 
$\mu$, and adopted the mass function described in \citet{Jung2022}. In the subsequent stage, we computed 
the lensing observables corresponding to the lens and source parameters utilizing the relationships in 
Eq.~(\ref{eq4}). Subsequently, we formed posterior distributions for $M$ and $\dl$ by assigning a weight 
to each synthetic event. This weight was determined by the expression:
\begin{equation}
w_i = \exp\left(-{\chi_i^2 \over 2} \right);\qquad
\chi_i^2 = 
\left[{t_{{\rm E},i}-\te \over \sigma(\te)} \right]^2 + 
\left[{\theta_{{\rm E},i}-\thetae \over \sigma^2(\thetae}\right]^2,
\label{eq5}
\end{equation}
where $(t_{{\rm E},i}, \theta_{{\rm E},i})$ represent the time scale and Einstein radius of each 
synthetic event, $(\te, \thetae)$ denote the measured values, and $[\sigma(\te), \sigma(\thetae)]$ 
are their uncertainties.

Figures~\ref{fig:eleven} and \ref{fig:twelve} depict the posterior probability distributions for the 
lens masses and distances associated with the events. Table~\ref{table:seven} provides a summary of the 
estimated masses of the host ($M_{\rm host}$) and planet ($M_{\rm planet}$), along with the distance 
and projected separation ($a_\perp$) between the planet and its host. For each parameter, the median 
value is presented as the central representative value, with the lower and upper limits determined as 
the 16\% and 84\% of the posterior distribution, respectively.

The identified planetary systems exhibit several common traits. Firstly, all planets orbit low-mass 
host stars, which are notably less massive than our Sun. The range of host star masses falls between 
$0.32$ and $0.58$ times the mass of the Sun. Secondly, the planets themselves are giants, exceeding 
the mass of Jupiter in our solar system within a range of $1.1$ to $10.7$ times the mass of Jupiter. 
Finally, all these planets are situated well beyond the ice line of their host stars, classifying 
them as ice giants.

Each panel of the posterior distributions separates the probability contributions from the disk and
bulge lens populations using blue and red curves, respectively. The combined contribution is shown
by the black curve. Table~\ref{table:seven} details the probabilities for the disk ($p_{\rm disk}$) 
and bulge ($p_{\rm bulge}$) populations.  Notably, for KMT-2022-BLG-0732, the lens is more likely 
situated within the disk  with a probability $p_{\rm disk} \sim 69\%$. Conversely, the remaining events 
exhibit a higher probability of residing in the bulge with probabilities $p_{\rm bulge} \gtrsim 64\%$.

\section{Summary and conclusion}\label{sec:six}

We conducted analyses of four microlensing events KMT-2020-BLG-0757, KMT-2022-BLG-0732,
KMT-2022-BLG-1787, and KMT-2022-BLG-1852, for which the light curves commonly exhibit positive 
deviations and subsequent negative deviations.  Unlike the usual brief anomalies observed in 
typical planetary microlensing events, the deviations in these events extend over a significant
portion of the light curves.  This prolonged deviation poses challenges in promptly identifying 
the presence of planets from the anomalies.

Our analysis revealed that each event's anomaly was caused by a planetary companion situated
within the Einstein ring of the primary star. The positive deviation of the anomaly was generated as
the source traversed one of the planetary caustics induced by a close planet, while the negative
deviation occurred as the source passed through the extended region of minor-image perturbations.

Upon estimating the physical parameters using the measured lensing observables of the events, we
found several common features among the identified planetary systems. First, all the planets
orbit low-mass host stars, significantly less massive than our Sun, ranging from 0.32 to 0.58 solar
masses. Second, these planets themselves are classified as giants, exceeding the mass of Jupiter in
our solar system, with masses ranging from 1.1 to 10.7 times Jupiter's mass. Finally, all the
planets reside well beyond the ice line of their host stars, making them ice giants.

\begin{acknowledgements}
Work by C.H. was supported by the grants of National Research Foundation of Korea 
(2019R1A2C2085965).
This research has made use of the KMTNet system operated by the Korea Astronomy and Space Science 
Institute (KASI) at three host sites of CTIO in Chile, SAAO in South Africa, and SSO in Australia. 
Data transfer from the host site to KASI was supported by the Korea Research Environment Open NETwork 
(KREONET).
This research was supported by the Korea Astronomy and Space Science Institute under the R\&D
program (Project No. 2023-1-832-03) supervised by the Ministry of Science and ICT.
The MOA project is supported by JSPS KAKENHI Grant Number JP24253004, JP26247023, JP23340064, 
JP15H00781, JP16H06287, JP17H02871 and JP22H00153.
J.C.Y., I.G.S., and S.J.C. acknowledge support from NSF Grant No. AST-2108414. 
Y.S.  acknowledges support from NSF Grant No. 2020740.
C.R. was supported by the Research fellowship of the Alexander von Humboldt Foundation.
This work was authored by employees of Caltech/IPAC under Contract No. 80GSFC21R0032 with the 
National Aeronautics and Space Administration.
V.B. is supported by PRIN 2022 CUP D53D23002590006.
R.F.J. acknowledges support for this project provided by ANID's Millennium Science Initiative 
through grant ICN12\textunderscore 009, awarded to the Millennium Institute of Astrophysics (MAS),
and by ANID's Basal project FB210003.
This project has received funding from the European Union's Horizon 2020 research and innovation 
program under grant agreement No. 101004719 (OPTICON - RadioNet Pilot). This work is supported 
by the Polish MNiSW grant DIR/WK/2018/12.
\end{acknowledgements}

\end{document}